\newcommand{\beq}{\begin{equation}}
\newcommand{\eeq}{\end{equation}}
\newcommand{\pa}{p^{(0)}}
\newcommand{\pb}{p^{(1)}}
\newcommand{\etal}{{\it et. al.}}
\newcommand{\Ninput}{N_d}
\newcommand{\omid}{o_{\rm mid}}
\newcommand{\FC}{{f_C}}
\newcommand{\tv}{{\tt t}}
\newcommand{\atv}{\bar{\tt t}}
\newcommand{\kk}{{\,k}}
\newcommand{\p}{{\tt p}}
\newcommand{\Prop}{{\cal P}}

\def\wisk#1{\ifmmode{#1}\else{$#1$}\fi}
\def\deg    {\wisk{^\circ}}
\def\ddeg    {\wisk{^\circ}}

\documentclass[12pt,preprint]{aastex}

\begin{document}

\title{Neural Networks as a tool for parameter estimation in 
        mega-pixel data sets}
\author{Nicholas G. Phillips}
\affil{Raytheon ITSS, Laboratory for Astronomy and Solar Physics, 
	Code 685, NASA/GSFC, Greenbelt, Maryland 20771}
\and
\author{A. Kogut}
\affil{Laboratory for Astronomy and Solar Physics, 
	Code 685, NASA/GSFC, Greenbelt, Maryland 20771}

\date{\today}

\begin{abstract}
We present a neural net algorithm for parameter estimation
in the context of large cosmological data sets.
Cosmological data sets present a particular challenge
to pattern-recognition algorithms
since the input patterns
(galaxy redshift surveys,
maps of cosmic microwave background anisotropy)
are not fixed templates overlaid with random noise,
but rather are random realizations
whose information content lies 
in the correlations between data points.
We train a ``committee'' of neural nets
to distinguish between Monte Carlo simulations
at fixed parameter values.
Sampling the trained networks
using additional Monte Carlo simulations
generated at intermediate parameter values
allows accurate interpolation
to parameter values for which the networks were never trained.
The Monte Carlo samples automatically provide
the probability distributions and truth tables
required for either a frequentist or Bayseian analysis
of the one observable sky.
We demonstrate that neural networks
provide unbiased parameter estimation
with comparable precision as maximum-likelihood algorithms
but significant computational savings.
In the context of CMB anisotropies, 
the computational cost for parameter
estimation via neural networks scales as $N^{3/2}$.
The results are insensitive to the noise levels
and sampling schemes typical of large cosmological data sets
and provide a desirable tool for the 
new generation of large, complex data sets.
\end{abstract}


%
%

\section{Introduction}


A fundamental question in cosmology is the origin and evolution  of large scale
structure in the universe.   The standard model for this evolution is the
gravitational growth and collapse  of initially small perturbations in the
primordial density distribution. This picture is supported by the detection of 
primordial Cosmic Microwave Background (CMB)
temperature anisotropies at a level of approximately one part in
$10^5$ by the Cosmic Background Explorer satellite and a series of ground-based
and balloon-borne experiments. 
Small perturbations on the matter and energy density in the early universe are
reflected in the temperature distribution of the CMB, providing a ``snapshot''
of conditions the early universe while the perturbations were still in the
linear regime.


One angular scale of particular interest is the horizon size at the  surface of
last scattering, the epoch when the universe cooled sufficiently to form
neutral hydrogen and allow the CMB photons to propagate freely.
Causally-connected regions at the surface of last scattering, as viewed from
the present epoch, subtend an angle
$$
\theta ~\sim ~1\ddeg7 ~\Omega_0^{1/2} ~( \frac{1100}{1 + z_{ls}} )^{1/2}
$$
\noindent
where $z_{ls}$ is the redshift at last scattering and $\Omega_0$ is the total
density of the universe relative to the critical (closure) density. Anisotropy
on scales larger than $\sim 2\deg$ ~reflect perturbations larger  than the
particle horizon and thus probe the primordial density distribution. On scales
smaller than 2\deg, causal mechanisms become important and modify the
primordial density in model-specific ways.

Oscillations
in the coupled photon-baryon fluid within the primordial potential wells  prior
to decoupling  lead to signature oscillations (``acoustic peaks'') in the
angular power spectrum of the CMB \cite{PeeblesYu70}. 
A precise determination of the power
spectrum in this regime can probe a wealth of information about the early
universe. For instance, the angular scale of the first acoustic peak depends
primarily on the angle subtended by the Jeans mass at 
last scattering \cite{HuSugiyama95}. It is
thus a direct probe of the large-scale geometry of the universe and hence the
density parameter $\Omega_0$. The width of the acoustic peak depends on the
sound speed at decoupling, and in turn on the baryon density and Hubble
constant as $\Omega_b h^2$. A measurement of this quantity could be compared
with similar determinations based on primordial 
nucleosynthesis \cite{Copi95,Olive00}. 
See {\it e.g.} \cite{Hu99}, for a review of how the different
physical processes in the
early universe leave their imprint on the CMB anisotropies.
Measurements
of the CMB anisotropy on medium angular scales thus offer an elegant
determination of such parameters as the
curvature, 
density,
matter density,
baryon density,
Hubble constant,
cosmological constant,
density perturbation index $n$,
scalar/tensor ratio,
and ionization history
of the universe.


From the one observable sky, we want to infer the values of these
cosmological parameters with minimal uncertainty in the shortest possible time.
Theoretical models do not predict a specific template for the CMB anisotropy (a
hot spot at this location, a cold spot over there), but rather predict a
statistical distribution usually expressed in terms of the angular power
spectrum. Deriving the power spectrum from the data (or more generally,
deriving model parameters directly from the sky maps) involves accounting for
angular correlations between pixels, precluding use of simple linear
least-squares techniques. The accepted standard in the CMB community has been
the generalization of least-squares techniques as implemented in maximum
likelihood algorithms (see, e.g.,
(G\'{o}rski et al.\ 1996;
Tegmark, Taylor, \& Heavens 1997;
Bond, Jaffe, \& Knox 1998;
Borrill 1999).


The simplest method of parameter estimation uses a
goodness-of-fit test to compare a set of observables  $y_i \pm \sigma_i$
measured at a set of positions $x_i$ to a theoretical model $\Gamma_i$. If we
have $N_d$ data points $y_i$ and $N_p$ parameters $p_j$, we define
\begin{equation}
\chi^2 = \sum_{i=1}^{N_d}
\left( 
\frac{ y_i - \Gamma_i }{\sigma_i} 
\right)^2,
\label{simple_chi}
\end{equation}
where 
\begin{equation}
\Gamma_i(x) = \sum_{j=1}^{N_p} p_j X_j(x_i)
\label{linear_eq}
\end{equation}
is function of the parameters $p$
and some fixed basis functions $X(x)$.
We obtain the ``best-fit'' parameter values
by minimizing $\chi^2$ with respect to the parameters,
\begin{equation}
\frac{\partial \chi^2}{\partial p_j} = 0
\label{min_chi_eq}
\end{equation}
for the $j^{\rm th}$ parameter $p_j$. The least-squares system in Eq.
\ref{min_chi_eq} has the solution
\begin{equation}
p_j = \sum_{k=1}^{N_p} ~({\bf A}^{-1})_{jk} B_j
\label{normal_eq}
\end{equation}
where
\begin{equation}
{\bf A}_{jk} = \sum_{i=1}^{N_d} 
~ \frac{ X_j(x_i) X_k(x_i) }{ \sigma_i^2 }
\end{equation}
is an $N_p \times N_p$ matrix, and
\begin{equation}
B_k = \sum_{i=1}^{N_d} 
~ \frac{ y_i X_k(x_i) }{ \sigma_i^2 }
\end{equation}
is a vector of length $N_p$. 


If the data points are not independent, this relatively simple calculation
becomes much more costly. Covariance between the observed data points can
result from instrumental artifacts (correlated noise, instrumental resolution,
oversampled data) or from correlations in the underlying signal (for instance,
measuring in real space  a signal whose components are independent in Fourier
space). Equation \ref{simple_chi} can be generalized  to include the effects of
covariance,
\begin{equation}
\chi^2 = \sum_{i=1}^{N_d} \sum_{j=1}^{N_d}
(y_i - \Gamma_i) ({\bf M}^{-1})_{ij} (y_j - \Gamma_j),
\label{chisq_def}
\end{equation}
where
\begin{equation}
{\bf M}_{ij} = \langle 
~(y_i - \langle \Gamma_i \rangle)
~(y_j - \langle \Gamma_j \rangle) ~\rangle
\label{covar_def}
\end{equation}
is the $N_d \times N_d$ covariance matrix  and the brackets denote an ensemble
average.

Conjugate gradient techniques can solve for $\chi^2$ without expliciting
solving for ${\bf M}^{-1}$ and thus avoiding the  $O(N_d^3)$ operations
this would incur. But if the covariance matrix ${\bf M}$ depends on the
parameters $p_j$, then minimizing $\chi^2$ will produce biased estimates
for $p_j$.
Maximum-likelihood parameter estimation provide a tool  to overcome 
this limitation.
For a multivariate Gaussian
distribution, the probability of obtaining the observed data $y_i$ given a set
of parameters $p_j$ is
\begin{equation}
{\cal L} = P(y | p) =
(2\pi)^{-N_d/2} ~
\frac{ \exp( -\frac{1}{2} \chi^2 )}
     { |{\bf M}|^{1/2} }
\label{like_def}
\end{equation}
where $\chi^2$ is defined in Eq. \ref{chisq_def}. The ``best'' choice of
parameters is that which maximizes the likelihood function ${\cal L}$. The
curvature of the likelihood surface about the maximum defines the uncertainty
in the fitted parameters,
\begin{equation}
\delta p_j \geq \sqrt{ ({\bf F}^{-1})_{jj} }
\label{err_def}
\end{equation}
where
\begin{equation}
{\bf F}_{ij} = \langle
\frac{ \partial^2 L }
     { \partial p_i \partial p_j } \rangle
\label{fisher_def}
\end{equation}
is the Fisher information matrix
(Kendall \& Stuart 1969)
and $L = -\log({\cal L})$
(see Bunn \& Sugiyama 1995;
Vogeley \& Szalay 1996;
Tegmark et al.\ 1997;
Bond et al.\ 1998).

The maximum likelihood estimator is unbiased and asymptotically approaches the
equality in Eq. \ref{err_def}. However, these advantages come at a steep price:
both the $\chi^2$ and the determinant calculation  in Eq. \ref{like_def} scale
as $O(N_d^3)$, making brute-force techniques computationally infeasible. 
For $N_d > 10^6$ the time required  is measured in years, 
even on the most powerful supercomputers. 
A number of authors have suggested ways around this problem. Karhunen-Lo\`eve
eigenvalue techniques produce moderate data compression, reducing the $N_d$
original data points to $N^\prime \approx N_d/10$ eigenmodes
(Bond 1994;
Bunn \& Sugiyama 1995;
Tegmark et al.\ 1997).
However,  estimating cosmological parameters from the smaller set of eigenmodes
still scales as $(N^\prime)^3$ operations,  making such techniques undesirable
for mega-pixel data sets.


Oh et al.\ (1998) derive a method for likelihood evaluation  using a
Newton-Raphson quadratic iteration scheme. This method utilizes  a conjugate
gradient algorithm to evaluate $\chi^2$.  The determinant is first
approximated  using azimuthal symmetry of the noise matrix (appropriate for
full-sky CMB maps), then corrected using Monte Carlo simulations. The method
provides a nearly minimum-variance estimate of the angular power spectrum for
CMB anisotropy maps in $O(N_d^2)$ operations and $O(N_d^{3/2})$ storage;
cosmological parameters can then be derived by comparing the power spectrum to
various models. Although this algorithm is fast enough for mega-pixel data
sets, it has several weaknesses. Since it is a Newton-Raphson iterative scheme,
it requires a sufficiently good starting estimate  to guarantee convergence to
true maximum.   It is also optimized to estimate the power spectrum, rather
than the underlying cosmological parameters -- when used as a root-finding
technique  in parameter space, the radius of convergence is small and the
problems associated with parameter covariance become severe.


The last few years have seen the development of a 
number of techniques designed to overcome the
$N_d^3$ problem, most of them focusing on estimating the power spectrum.
These techniques include Szapudi \etal\ (2000), who have developed a method
based on using the two-point correlation function. Dor\'e \etal\ (2001)
relies on an hierarchical implementation of the usual quadratic estimator
to the power spectrum. Both methods are known to scale as $N_d^2$, with the
first hoping to be improved to $N_d\log N_d$ while the second may scale
as $N_d$ (with a large prefactor). There is also the Monte Carlo estimator
to the power spectrum developed by Hivon \etal\ 2001, which has been used
\cite{Netterfield01} to analysis the BOOMERANG data \cite{deBernadis00}.
All these approximate techniques rely on first estimating the power spectrum
and then using this estimated power spectrum to determine the most likely
cosmological parameters. Douspis \etal\ (2001) have shown that the data sets
are not fully described by just their band power sets.

Borrill (1998) offers a global solution to bound the likelihood. This method
uses Gaussian quadrature to bound the likelihood at {\it any} point in
parameter space  (not just near the likelihood maximum); it is thus well
suited  to search parameter space using the minimum-variance direct pixel
basis. However, the method requires $O(N_d^{7/3})$ operations for each
likelihood evaluation and is thus significantly slower than the method of Oh et
al.\ (1998). More importantly, it can only provide bounds on the likelihood,
fixing $\log({\cal L})$ to accuracy of a few percent. Since $\log({\cal L}) >
N_d$ (a large number), errors of a few percent can create significant bias in
the location of the likelihood maximum. 

The Microwave Anisotropy Probe (MAP),
launched in the summer of 2001, will measure the full sky in 5 frequency
bands  with over a million pixels per band \cite{Bennett95}. The Planck
Surveyor mission, scheduled later in the decade, will produce maps with over
$10^7$ pixels \cite{Bersanelli96}. Maps of such size can not be analyzed
with brute-force maximum likelihood techniques. 

Figure \ref{fitting-model}
 shows typical information flow for cosmological parameter estimation.
We compare data (CMB sky maps) to a parameterized model using some mathematical
``machinery'' to derive a set of parameters describing the data.  Although
least squares methods and their generalization to maximum-likelihood techniques
are common choices for this machinery, they are not the only choices possible.
For mega-pixel data sets, these deterministic methods are computationally
infeasible.  Neural net algorithms provide an alternative machinery
for parameter estimation in large, complex data sets.

Neural networks have been used previously in astronomy 
for galaxy classification \cite{Lahav96,Andreon00} and
periodicity analysis of unevenly sampled data as applied to
stellar light curves \cite{Tagliaferri00}.
They have also been used to analyze
stellar spectra \cite{BailerJones97,BailerJones98,BailerJones00},
with results comparable to traditional methods. 
However, Bailer-Jones {\it et. al.}
compared data to a deterministic model (stellar spectra),
whereas cosmological applications examine random patterns
drawn from parameterized stochastic models.
We demonstrate the
generality of our algorithm by considering different problems with
the same network architecture.


We use neural networks as
a complementary approach to  cosmological parameter estimation
in large data sets.
They are a forward algorithm in that they  ``learn'' the
differences between two different parameter data points 
by being trained on sets of simulated data sets drawn at the
specified parameter points.
We find the computational cost for training the network,
in the context of CMB anisotropy,
requires $O(N^{3/2})$ operations and thus provides a substantial improvement
over brute-force maximum-likelihood methods.

Neural networks do not require that we
specify one single statistic of {\it a priori} interest.
As the network is trained, it determines how it will discriminate.
The information required to separate different parameter points
comes from the training set simulations.
After selecting a pair of parameter points, a committee of networks
are trained on simulations taken at each of the parameter points.
We interpolate by sampling
the trained networks with simulated data drawn at parameter points between
the training sets. 
This takes advantage of the fact that for large input patterns,
the number of sampling sets will be much less than the number of
training passes.
The algorithm naturally provides the
distributions necessary to understand the confidence levels of the parameter
fit when the physical data is analyzed. This sort of sampling of trained
networks becomes a key component in utilizing a Bayesian approach to
parameter estimation, see \cite{Christensen00,Rocha00} for discussions of
the Bayesian approach in the context CMB anisotropies and
\cite{MacKay95}, along with \cite{Bishop95}, for neural networks.

%


We present a brief review of the neural network algorithm and
discuss how this model is used for parameter estimation. To demonstrate
the feasibility of this algorithm, we show how we can estimate parameters
in three different examples. The same neural network and parameter estimation
algorithm is used in all the examples.
Using these examples, we vary the size and noise characteristics
of the simulated data sets to determine the computational scaling.

\section{Neural Network Algorithms}

Parameter estimation can be thought of as an
exercise in pattern recognition,
for which an extensive literature exists
(see, e.g.,  
  McCulloch and Pitts 1943; 
  Rosenblatt 1962; 
  Watanabe 1969;
  Hopfield 1982; 
  Rumelhart, Hinton, \& McClellend 1986;
  Grossberg 1988).
Neural networks, an implementation of parallel distributed processing 
architecture, are well suited to pattern recognition.
Although not widely recognized as such, neural networks are also well suited
for the problem of parameter estimation in large, complex
data sets.

\begin{figure}
\plotone{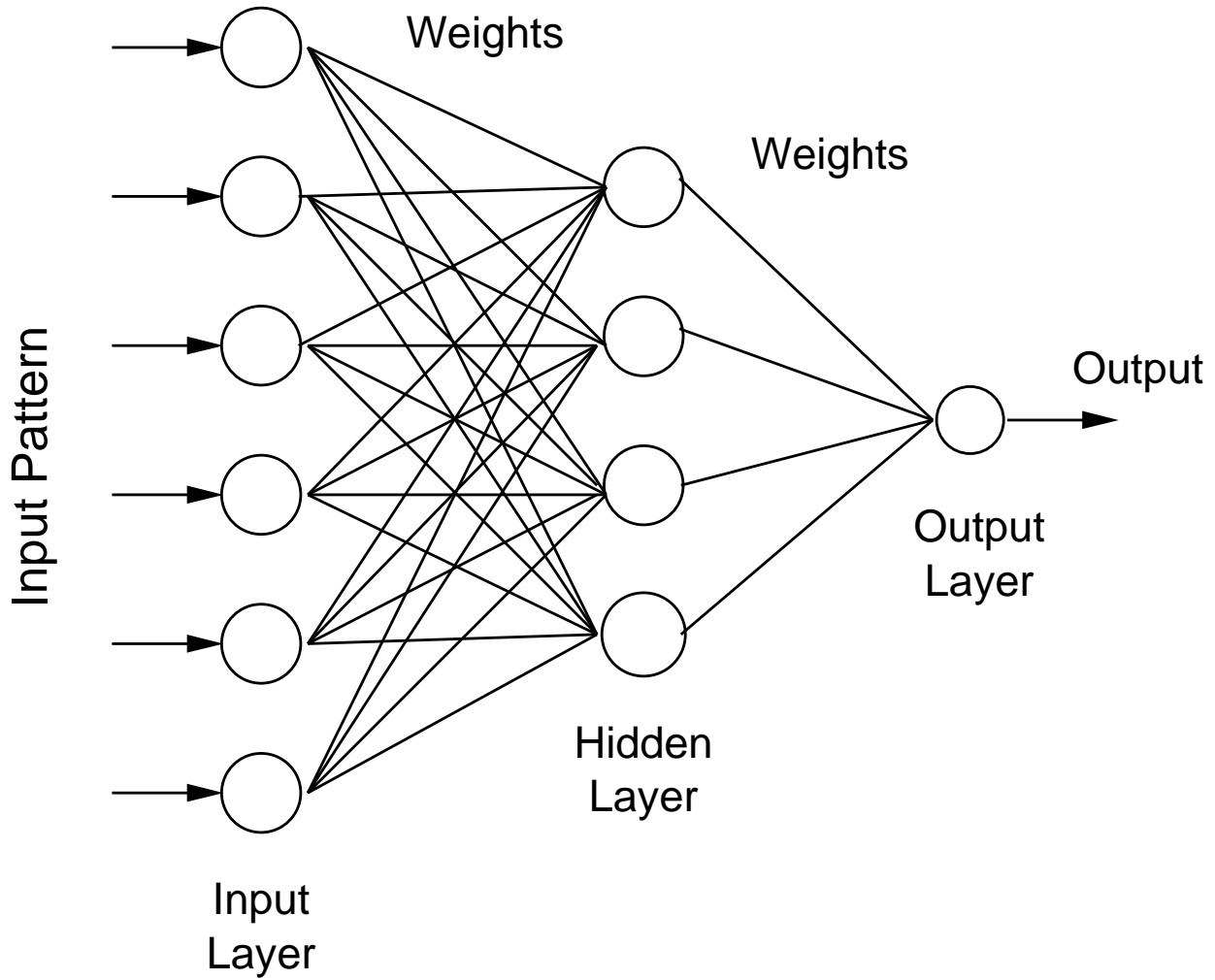}
\caption{
Neural network architecture used for parameter estimation. The data (noisy sine
wave, CMB maps or galaxy survey)  are the input pattern and the neural network
feeds forward through the hidden layer to the output unit.  The output value is
used to derive the probability distributions for the samples and estimated
parameter for the observed data set. Typically there are fewer hidden units
than input units. 
}
\label{skymap-neural-net}
\end{figure}

The basic building block of a neural network is the neuron, consisting of any
number of inputs and a single output. A neuron sums its inputs to determine its
output via some (non-linear) transfer function. A neural network 
consists of layers
of neurons connected together via matrices of weights (Figure
\ref{skymap-neural-net}). A network is trained to produce a desired output by
repeatedly presenting the network with a set of known input patterns, then
adjusting the weights until the network produces the desired output for each
known input. When the network is later presented with an unknown input, the
output will reflect which of the known inputs  the unknown input most closely
resembles. 

Figure \ref{skymap-neural-net} shows a typical network architecture. The first
layer in a network is its input layer,  whose outputs reflect the pattern
presented to the network. Any number of hidden layers lie between the input and
output layers,  from which the results of the network are read. We use one
hidden layer and a single output unit.  We thus have one matrix of weights
connecting the input pattern to the inputs of the hidden layer and another
matrix connecting the outputs of the hidden layer to the inputs of the output
unit.

The information content of a neural network lies in the matrices of weights
connecting layers
(analogous to the information content of the covariance matrix for maximum
likelihood algorithms). The weights are set during the training process, in
which we present the network with a series of known inputs and adjust the
weights to obtain the desired outputs. For a network with $\Ninput$ input
units, we let  ${\bf X}_{\rm patt}$, a $\Ninput$ element vector, represent an
input pattern. The output of the network can be viewed as a function of the
input pattern: $o = o({\bf X}_{\rm patt})$. For a given pattern  we associate
an output target $t_{\rm patt}$ (in our case a single value, but for a more
general network, this would be a $N_{\rm output}$ vector). A training algorithm
is a method for adjusting the weights to minimize the total error $E =
\sum_{\rm patt} E_{\rm patt}$, where
\beq
	E_{\rm patt} = \frac{1}{2}
        \left(o({\bf X}_{\rm patt}) - t_{\rm patt}\right)^2
\eeq
We use the back propagation method  (Rumelhart, D.E., Hinton, G.E.,  and
Williams, R.J.\ 1986). According to this method, we present a series of
$N_{\rm Train}$ patterns to the network, one at a time. For each pattern,
we compare the output with the associated target
and determine the error. We
use this error to adjust the weights  connecting the hidden layer to the output
layer, and then back-propagate it to the weights between the hidden and input
layers.  Note that the training depends on the difference between the desired
and actual outputs - no {\it a priori} statistical test 
for the inputs is specified.


Neural nets achieve computational savings by  training with simulated data
sets, as opposed to inverting a large matrix. The computational cost to train a
neural net 
is found to be
\begin{equation}
N_{\rm CPU} \propto \Ninput^\alpha,
\label{computation_eq}
\end{equation}
where the exponent $\alpha$ depends on the particular problem. There are two
main contributions to this cost: the array multiplication needed to sum the
weights to evaluate a single pattern, times the number of training patterns
required. In general we find $1 \le \alpha < 3$, depending on the specific
problem. This is never worse than standard maximum likelihood methods, and at
best the computational cost grows linearly with the input pattern size. For the
specific case of parameter estimation in mega-pixel CMB maps, the computational
cost scales as $N_{\rm CPU} \propto \Ninput^{1.5}$. Neural networks provide
significant computational savings over both standard maximum-likelihood
algorithms and the $N^2$ scaling for the fastest known approximate method.

\section{Neural Networks and Parameter Estimation}

Neural networks can be trained to estimate the value of a continuous parameter,
and can reliably interpolate to parameter values intermediate between the 
training values.  Though this idea is not new to astrophysics, {\it e.g.}
\cite{BailerJones97,BailerJones98,BailerJones00}, our method is fundamentally
different from that of Bailer-Jones {\it et. al.} in that our patterns are
random samples drawn from a parameterized parent population. The Bailer-Jones
{\it et. al.} method is a matter of template matching; randomness only enters
their input patterns as instrument noise. 
For CMB and other cosmological data, the patterns themselves are
intrinsically random. Nonetheless, using the same basic
neural network architecture, we can train the networks to discriminate
stochastic patterns that differ according to the parent population from which
they are drawn.

\begin{figure}
\plotone{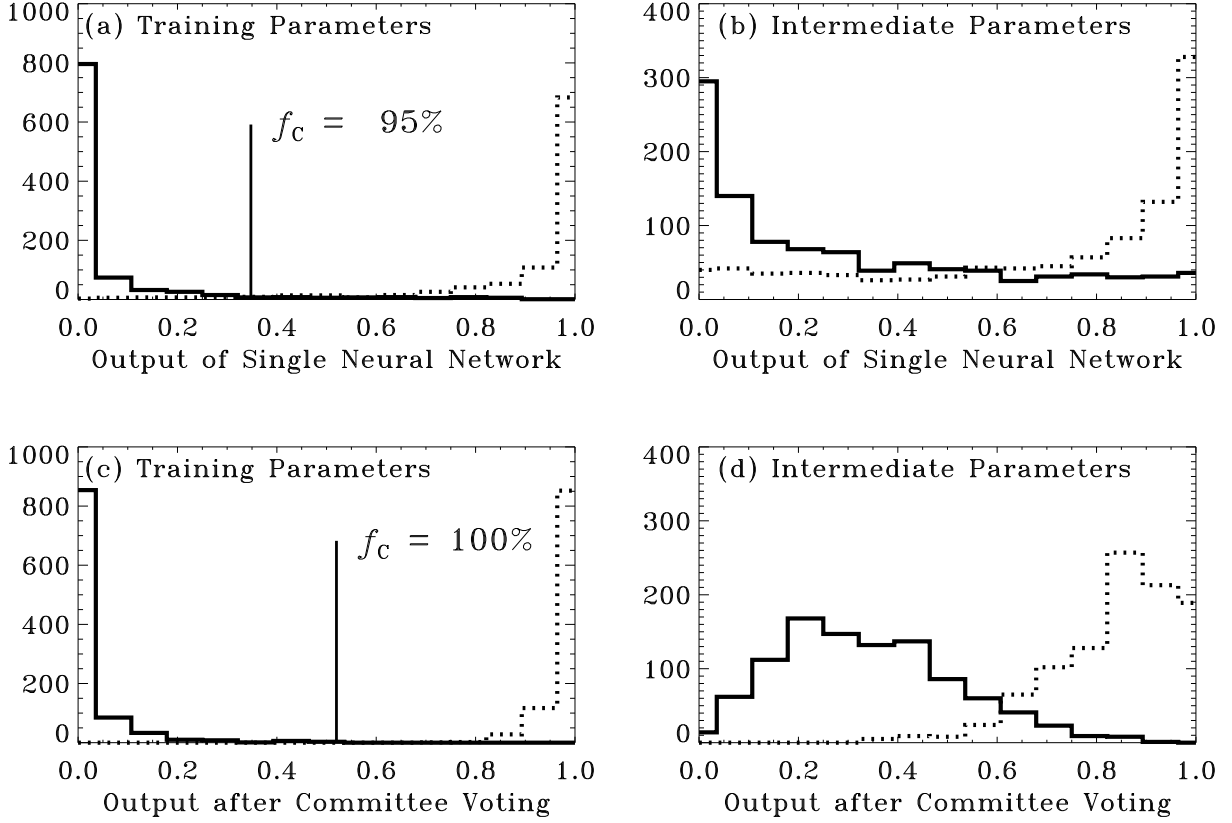}
\caption{
Sample neural network output distributions.
(a) Solid line is the distribution of output values for an independent set of
samples drawn at the same parameter value $\pa$ for which the network was
trained to the target $0$. The dotted line is for samples drawn
at $\pb$, the value for the target $1$. The vertical line is
the midpoint value $\omid$ that maximizes the fraction correct $\FC$, 
{\it i.e.}, all output values $<\omid$ are identified as being drawn at 
$\pa$ while the rest at $\pb$. 
(b) The output of the same network, but for samples drawn at parameters 
intermediate to $\pa$ and $\pb$. The solid line is for $p$ close to $\pa$ 
and the dotted line for a choice close to $\pb$. 
(c) Solid line is the average truth value $\atv$ for the $\pa$ patterns, 
averaged over a committee of 50 networks, the only difference between
the networks being the initial randomization of the weights.
The dotted line is for $\pb$.
(d) The average truth value for the same sets of samples in (b).
The averaging has produced two well defined 
peaks that are cleanly separated.
Distributions of $\atv$ like those in (c) and (d) 
become the basis for predicting which estimated parameters to 
associate with the average truth values, the ouput due to presenting a 
committe of networks with an unknown pattern.
}
\label{sample-outputs}
\end{figure}

We start by training a network to differentiate between  simulated data sets
(including instrument noise and other artifacts) generated at a pair of
discrete parameter values. The back-propagation adjusts the weights until the
network outputs target value 0 when presented with the first set of patterns,
and target value 1 when presented with the second. In practice, since the
information distinguishing different parameter values is in the correlations,
not the actual pixel values, any single network will not train to a sufficient
degree. We improve the situation by training a small committee of networks and
polling them to get a consensus opinion. Now we find simulations generated
at the training parameter values produce two well defined peaks.
Simulations generated with an
intermediate parameter value (never present in training data) yield outputs
peaking somewhere in between, depending on whether the new parameter value is
closer to the first or second  training value (see Fig \ref{sample-outputs}).

We quantify this behavior by presenting the trained network with a set of new
inputs drawn from a grid of intermediate parameter values, and derive for each
intermediate parameter value the corresponding probability distribution of
output values. When the networks are later presented with an  unknown pattern,
each distribution gives the probability that the unknown  pattern was generated
with a parameter value corresponding to that grid point. The interpolated
parameter is the probability-weighted mean. The grid samples  all use the same
trained networks; the sampling of the networks at different  grid points is
faster than the training since we usually need many more training sets than
sampling sets, thus no great  computational cost is incurred. 
Although we focus below on estimating a single parameter, 
the method is readily extended to multi-parameter fits.

The simulations are viewed as random variables ${\bf X}(p)$ taken from an
underlying parameterized probability distribution, with $p$ the parameter.
Assuming the data lies
between two extreme values $\pa$ and $\pb$ we train a network to the target
$t^{(0)}=0$ for realizations ${\bf X}(\pa)$ and $t^{(1)}=1$ for ${\bf X}(\pb)$
by repeatedly presenting the network with new samples at $\pa$ and $\pb$ and
back propagating the error.  Once trained, the network will give an output
$o\left({\bf X}\right)$ between 0 and 1 for any input parameter value. 
If $p < \pa$, the output clips at 0, while if $p > \pb$ the output clips at 1. 

Once we have trained a network, we can present additional, statistically
independent samples drawn at $\pa$ and $\pb$. Figure \ref{sample-outputs}a
shows the output distributions for $1000$ patterns of each parameter. 
We see the network has successfully trained in that the $\pa$ distribution
is peaked at $o=0$ while the $\pb$ samples at $o=1$.
To be able to interpolate to intermediate values, we will need to present
samples drawn at intermediate parameter values. Fig \ref{sample-outputs}b
shows the output distributions for an additional $1000$ samples each for two
parameters, one just a little larger than $\pa$ and one a little smaller than
$\pb$. These distribution also show the same tendency to peak close to the
limits of $0$ and $1$, but not as strongly as those drawn at the trained
parameters. In effect, the network is chosing which of the training parameters
these new patterns, for which it was never trained, most closely resemble.
At it stands now, this tendency makes it hard to construct the probability
distributions we need for parameter estimation.

By using a committee of networks, we take advantage of this peaking tendency.
We want to determine the committee consensus and from this get the distributions
we seek. The first step is converting the continuous output value into a discrete
{\it truth values} $0$ or $1$.
For each trained network,
we associate a midpoint value $\omid$ and for any input pattern ${\bf X}$,
we define its truth value according to 
\beq
\tv \equiv \tv\left({\bf X}\right) = \left\{
\begin{array}{cc}
  0 ; & o\left({\bf X}\right) \le \omid \\
  1 ; & o\left({\bf X}\right)  >  \omid
\end{array}
\right..
\eeq
We interpret $\tv({\bf X}) = 0$ as indicating the pattern was drawn from the parent
population with parameter $\pa$, and similarly we associate $\tv=1$ with $\pb$.

To determine $\omid$, we present $N$ samples drawn at $\pa$ and 
at $\pb$. For each of these sets and any $\tilde\omid$, we obtain the
truth values $\tv_i^{(0)}$ and $\tv_i^{(1)}$, $i=1,\ldots,N$.
With this, ${\tt n}^{(0)} = \sum_i(1-\tv^{(0)}_i)$ is the number patterns drawn
at $\pa$ correctly identified as drawn at $\pa$ and 
${\tt n}^{(1)} = \sum_i \tv^{(1)}_i$ similarly at $\pb$. We chose $\omid$ to
maximize $\FC = \frac{1}{2N}({\tt n}^{(0)} + {\tt n}^{(1)})$ and refer
to $\FC$ as the {\it fraction correct}, our main measure of how well a network
has trained. In Fig \ref{sample-outputs}a, $\omid$ is marked with the vertical
line and we find $\FC=94\%$.

We note $\FC$ allows us to determine the optimal number
of training passes, $N_{\rm Train}$.
Starting with an initially randomized
network, as the network trains, we intermittently pause the training and
sample the network to determine $\FC$. It steadily increases to a maximum value 
and then levels out: the minimum for the training error $E$ has been reached. 
We take $N_{\rm Train}$ to be just where this plateau starts.

For each network, any given input pattern is converted into discrete
truth values $\tv$. We now form a committee of such networks, where the
only difference between the networks is the initial randomization of the
weights. We find committee sizes $N_{\rm net}\sim 50$ sufficient. After
presenting any given pattern ${\bf X}$ to the committee, we have the
collection of truth values $\tv({\bf X})_m$, $m=1,\ldots,N_{\rm net}$.
We view each $\tv_m$ as the vote from network $m$ as to whether the
pattern resembled those drawn at $\pa$ or $\pb$. The committee consensus
is formed by generating the {\it average truth value}
\beq
\atv({\bf X}) = \frac{1}{N_{\rm net}}\sum_{m=1}^{N_{\rm net}}\; \tv({\bf X})_m.
\eeq
Figure \ref{sample-outputs}c shows the distribution of average truth values
for the same set of samples drawn at $\pa$ and $\pb$ used in
Fig \ref{sample-outputs}a. They are now even more sharply peaked about
$\atv=0$ and $\atv=1$ and in terms of the average truth value, $\FC=100\%$.
More important are the distributions displayed in Fig \ref{sample-outputs}d.
These are for the same intermediate samples as used in Fig \ref{sample-outputs}b;
we have now two well defined distributions with peaks intermediate to the
peaks for the $\pa$ and $\pb$ samples. This is the general trend 
when we work in terms of the average truth value:
as the parameter $p$ is swept
from $\pa$ to $\pb$, we get well defined distributions whose peak moves from
$\atv\sim 0$ to $\atv\sim 1$.

We utilize this behavior to have our 
networks interpolate parameter
values for which it was {\em never} trained. We need a function that
maps an average truth value to an estimated parameter: $\atv\mapsto \p(\atv)$.
We built this function by sampling the committee of networks with samples
drawn at parameter values intermediate to the training values (computationally 
a small cost compared with training a network). These in turn are used to 
determine the optimal function $\p(\atv)$. The results of such sampling also
automatically provide the distributions needed for either a frequentist's or
Bayesian analysis. 
Selecting $K$ parameter values
uniformly distributed between the training values,
$p^\kk = \pa,\;\pa + \Delta p,\;\pa + 2\Delta p,\;\ldots,\pb$, we
generate $N$ samples at each of these parameter values.
The samples are presented to the committee of networks and thus
for each sample ${\bf X}_i$ drawn at each parameter value $p^\kk$, its
average truth value $\atv_i^k$ is computed.
We determine the
parameter estimation function $\p(\atv)$ by minimizing 
the distances between estimated and true parameter values,
{\it i.e.} the error
\beq
E_\p = \frac{1}{2} \sum_k \sum_i \left( \p(\atv_i^k) - p^\kk\right)^2.
\eeq
The $\atv$'s will take on discrete values,
$\atv = 0,1/N_{\rm net},2/N_{\rm net},\ldots,1$, and thus so
will the parameter estimation function:
$\p(\atv) = \p_j$ when $\atv = j/N_{\rm net}$. Let $n^k_j$ be the number of
samples drawn at $p^\kk$ with $\atv^k_i = j/N_{\rm net}$. Then we can re-write our
error as 
$E_\p = \frac{1}{2} \sum_{k,j} n^k_j\left( \p_j - p^\kk\right)^2$.
Minimizing this yields
\beq
\frac{\partial E_\p}{\partial \p_j} = 0 \quad \Rightarrow \quad
\p_j = \frac{\sum_k n^k_j \, p^\kk}{\sum_k n^k_j},
\label{eq-pfct-define}
\eeq
{\it i.e.}, each binned value of the parameter estimation function is the
parameter weighted average of the number of patterns of each sample that
fall into that bin. Figures \ref{sample-outputs}c and \ref{sample-outputs}d
are examples the $n^k_j$ distributions.

Given an observed data set, represented as the input pattern ${\bf X}_{\rm obs}$,
we present it to the committee of networks and obtain its average truth value
$\atv_{\rm obs}$. From this, we get the estimated parameter value
$\p_{\rm obs}$. We may now generate a set of samples at this
parameter value, $\left\{{\bf X}_i(\p_{\rm obs})\right\}$, and estimate their
parameters, $\p_{i,\rm fit}$.
The distribution of the $\p_{i,\rm fit}$'s 
provides a self-consistent method for converging to the final estimation of the
parameter. We begin with a wide separation between training values in an
attempt to bracket the unknown input. We know the initial range is  too narrow
if either $\p_{\rm obs} \sim \pa$ or $\sim\pb$,  that is, if the output clips at
either end of its trained range. The distribution of the $\p_{i,\rm fit}$'s
 also tells
us when we have converged on final answer.  We compare the width in the fitted
parameter distribution to the separation between training values, and select
new (closer) values. We interate this refinement of $\pa$ and $\pb$ until the
mean $\bar \p_{\rm fit}$ doesn't change or the width of the fitted distribution 
matches the separation of training parameters. In practice convergence
typically requires only two or three iterations.

We illustrate how we determine the confidence levels of our parameter estimation
via a Bayesian analysis of the results. This is not the only way to proceed;
if one preferred, a frequentist's approach could be taken, similar to the above
paragraph. From the Bayesian viewpoint, we are interested in posterior
distribution $\Prop(p|\p_{\rm obs})$: given the observed/estimated
parameter $\p_{\rm obs}$, what is the probability the true parameter is $p$. 
With this, we view the uncertainty in the estimated parameter as
\beq
\sigma^2(\p_{\rm obs}) =
\int p^2\, \Prop(p|\p_{\rm obs}) dp
- \left(\int p\, \Prop(p|\p_{\rm obs}) dp\right)^2,
\eeq
the width of true parameter values around $p_{\rm obs}$ that have a reasonable
probability of being identified as $p_{\rm obs}$. We can also use this determine
the confidence levels. We use Bayes Theorem to express
this in terms of the prior distributions:
\beq
\Prop(p|\p_{\rm obs}) = \Prop(\p_{\rm obs}|p)\frac{\Prop(p)}{\Prop(\p_{\rm obs})}.
\label{eq-bayes-thm}
\eeq
All these priors are readily available in the analysis we have developed so far.
If there are the same number of patterns in each of the $K$ sample sets, 
then we have a uniform prior on the parameters $p^\kk$ we sampled:
$\Prop(p^\kk) = 1/K$.
The probability of getting any particular $\p_{\rm obs}$ is proportional to the
total number of $\atv$'s that have the corresponding value and hence
$\Prop(\p_{\rm obs}) = \frac{1}{K N}\sum_k n^\kk_{\rm obs}$.
Given the true input parameter value is $p^\kk$,
the probability that it is identified as $\p_{\rm obs}$ 
is proportional to the number of samples with 
an average truth value that corresponds to $\p_{\rm obs}$: 
$\Prop(\p_{\rm obs}|p^\kk) = n^\kk_{j=\rm obs}/N$.
Thus, the probability distribution, Eqn. \ref{eq-bayes-thm}, is
$\Prop(p^\kk|\p_{\rm obs}) = n_{\rm obs}^\kk/\sum_k n_{\rm obs}^\kk$. 
The resulting expression for the
mean true parameter is the same as Eqn. \ref{eq-pfct-define}, 
our optimal choice for
defining the parameter estimation function. The 68\% confidence width for our
parameter estimation becomes
\beq
\sigma^2(\p_{\rm obs}) =  
	\frac{\sum_{\,k} (p^\kk)^2 \; n_{\rm obs}^\kk}
         {\sum_{\,k} n_{\rm obs}^\kk} 
   - (\p_{\rm obs})^2
\eeq
It is important to note all the information used to determine this uncertainty
in the fit was already generated during the determination of $\p(\atv)$.

We can use this Bayesian analysis to supplement our convergence criterion
outlined above. The posterior probability $\Prop(p|\p_{\rm obs})$ is zero
for $p<\pa$ or $p>\pb$, since there are no samples drawn outside the
training range. Thus the uncertainty will be artificially
suppressed close the to these limits. If $\p_{\rm obs}$ is within
$\sigma(\p_{\rm obs})$ of either $\pa$ or $\pb$, we need to select
new training limits, {\it i.e.}, we have encountered the clipping
previously discussed. If we are clear of either training limit and
$\sigma(\p_{\rm obs})$ is much less than the width of the training
interval, we should choose new, closer, training limits.

Whether we chose to do a frequentist's or Bayesian analysis, this method
of using neural networks for parameter estimation
benefits from not requiring the {\it a priori}
definition of a statistic or goodness-of-fit function. 
We only need to be able to simulate the model, often a computational
inexpensive task. Neural networks are ideally suited to working with
models where the crucial information is in the phase information,
{\it e.g.} topology of the Universe or tests for non-Gaussianity,
along with being a faster algorithm for conventional parameter
estimation problems.

\section{Examples}

To illustrate this algorithm, we analyze three examples. The first is the
estimation of the frequency of an irregularly sampled noisy sine wave. This
serves as a straightforward example of our method and shows how the  neural
network can recover a parameter value at which it was never trained. Next, we
fit the spectral index for CMB anisotropies based on a Gaussian model,
an example of the type of problem that will need accurate methods with low
computational cost. The final example is parameter estimation in the context of
large scale structure surveys, another arena with large data sets. This example
illustrates the strength of the neural network to not need an {\it a priori}
statistic.


\subsection{Irregularly Sampled Noisy Sine Wave}

\begin{figure}
\plotone{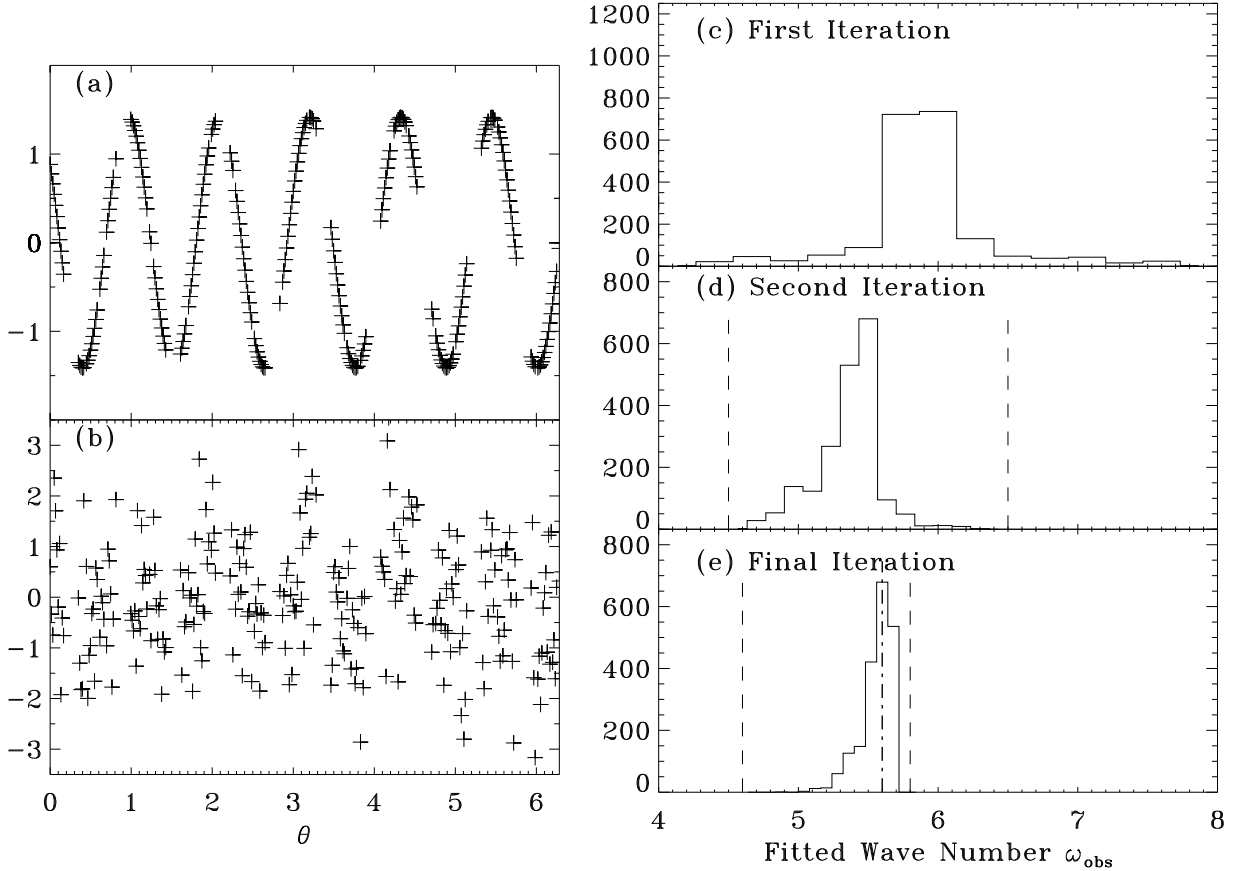}
\caption{
Neural net results from noisy time series.
 (a) Sample sine wave for $\omega_{\rm in}=5.60$, including 10 regularly
spaced gaps in $\theta$, along with 24 randomly placed gaps.
(b) Same sample sine wave, but now including Gaussian noise so that
the resulting input network pattern has a S/N=0.4
(c) Histogram of fitted $\omega_{\rm obs}$ for an initial network
training range of $\omega^{(0)}=4.0$ and $\omega^{(1)}=8.0$.
(d) Fitted values after second iteration.
Note that the training values now bracket the distribution
returned by the first iteration. 
Vertical dashed lines show the training frequencies.
 (e) Fitted values after third (final) iteration.
The width of the distribution matches the separation of training values.
Note that we correctly recover the input frequency,
$\bar\omega_{\rm obs}=5.56\pm 0.11$,
even though none of the networks were trained at this value.
}
\label{sine_wave}
\end{figure}

A common problem in astrophysics is the estimation of periodic signals in
noisy, irregularly sampled data sets. Observing limits often produce data
cutouts in time series; the resulting aliasing of power then precludes simple
Fourier transforms. Although other techniques such as periodograms exist for
this problem, time series provide a simple introduction to the capabilities of
neural nets. Our first example uses a time series consisting of  $\Ninput =
256$ points,
\beq
X_i(\omega) = \sqrt{2} A_s \sin\left(2\pi\omega\,t_i + \phi\right) + n_i
\eeq
where $t_i$ ranges from 0 to $t_f$. We restrict data coverage by including 10
uniformly spaced gaps of 10 pixels each, plus an additional 24 randomly-placed
gaps (Fig \ref{sine_wave}a).  $\phi$ is a  random phase and $n_i$ is Gaussian
random noise of zero mean and unit variance. The factor of $\sqrt{2}$ gives
signal-to-noise, $S/N$, of $A_s$ for each observation (Fig \ref{sine_wave}b).

This sine wave model is somewhere between a template and a random pattern.
If $\phi = {\rm const.}$, then this would be a matter of template matching, 
since but for the noise, the values of each of the pixels would be completely
determined for each frequency. This is akin to the classification of
stellar spectra \cite{BailerJones97,BailerJones98,BailerJones00}.
We are allowing $\phi$ to be a random variable and thus the
location the peaks and vallleys of the sine wave vary from pattern to pattern. 
Our model
has characteristics of a random pattern: the information about the frequency
is now in the correlations of the pixel values.

We train 50 networks using initial parameter range  $\omega^{(0)}=4.0$ and
$\omega^{(1)}=8.0$ to produce an output value of 0 for
noisy sine waves with $\omega=4.0$  and output value 1 for $\omega=8.0$. 
We then sample the trained networks at intermediate values  of
$\omega$ to generate nine probability distributions ${\cal P}(o;\omega_i)$,
$\omega_i=4.0,4.5,\ldots,8.0$.  Finally, we present the networks with 1000
realizations at frequency $\omega_{\rm in}=5.60$.  Figure \ref{sine_wave}c 
shows the distribution of recovered $\omega_{\rm obs}$ from these realizations.
The mean value is intermediate between the training inputs, with no output
close to the training values of $4.0$ or $8.0$. The distribution shows most of
the estimated parameters lie between $\omega^{(0)}=4.5$ and $\omega^{(1)}=6.5$,
suggesting we repeat the analysis with these new training values. Fig.
\ref{sine_wave}d shows the result after a second iteration. The peak of the
distribution of estimated parameter values has shifted somewhat from from the
previous iteration. Selecting new training values $\omega^{(0)}=4.6$ and 
$\omega^{(1)}=5.8$ to encompass the bulk of the estimated parameters, we do one
more iteration.
The peak for this distribution of estimated parameters has not shifted from the
previous set of networks and the  width of the distribution matches the range
of sampled frequencies. The distribution peaks at $\omega=\omega_{\rm in}$,
with fitted mean $\bar\omega_{\rm obs}=5.56\pm 0.11$.

Note that we correctly recover $\omega_{\rm in}$ despite never training any
network at any stage to  this value. Neural networks provide an unbiased
estimator for a continuous parameter without requiring prior knowledge of the
parameter value.


\subsection{Microwave Background Anisotropy}
Neural networks can also be applied to more challenging computational tasks.
Deriving cosmological parameters from maps of the cosmic microwave background
usually involves maximum likelihood algorithm whose computational cost ($N^3$)
makes them prohibitive for mega-pixel maps. The same neural net topology used
for our time series example provides accurate, rapid parameter estimation for
CMB maps as well.

On angular scales $\theta > 2\deg$, anisotropy in the cosmic microwave
background   corresponds to primordial density perturbations with scale-free
power spectrum  $P_k \propto k^n$, where $k$ is the wavenumber  and $n$ is the
power-law index.  We simulate the COBE-DMR full-sky maps of the CMB anisotropy
parameterized by the index $n$ \cite{Bond87}.  We smooth full-sky maps with a
Gaussian profile with a FWHM of $7^\deg$ to include the effect of the 
radiometer horn profile and add Gaussian noise with an variance of $(20~ {\rm
mK})^2/N_{{\rm obs},i}$ to each pixel to account for the  instrument noise
\cite{Bennett96}.

Foreground emission from our Galaxy dominates  the COBE data near the Galactic
plane, rendering it unusable for cosmological analyses. We use the galaxy cut
template of \cite{Banday97} to excise pixels with significant Galactic
emission. The cut sky represents an additional challenge for standard
maximum-likelihood analyses. In the absence of this cut, the data sets
represent full-sky coverage and can be decomposed in terms of orthogonal
spherical harmonics.  The resulting coefficients yield the power spectrum of
the CMB and hence the spectral index $n$. Once the galaxy cut is imposed, the
spherical harmonic functions are no longer orthogonal on the remaining pixels.
Any attempt to to obtain a harmonic expansion will result in the aliasing of
power between modes and an inaccurate power spectrum. Though a new orthogonal
set of basis functions can be computed for the cut sky (G\'{o}rski, 1994), this
is an $N^3$ problem as well.  Since neural networks estimate cosmological
parameters  using the real-space pixel values, they need not take the detour
through the power spectrum, and do not suffer aliasing of power. We simply
impose the cut for galactic emission and train each network using only the
remaining high-latitude pixels. As the network is trained ,
it automatically learns the effect of cuts in
the data, without requiring any symmetries in the cut data  (see {\it e.g.}
\cite{Oh99}).

\begin{figure}
\plotone{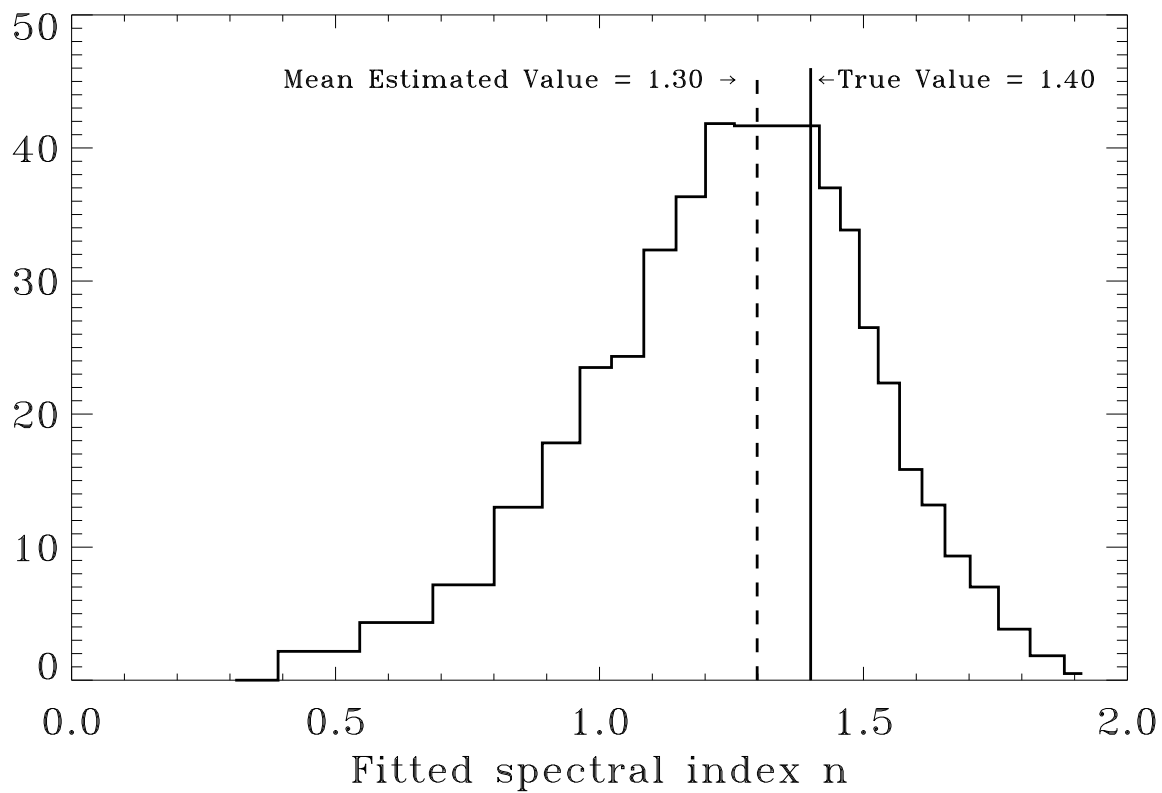}
\caption{
Fitted spectral index $n_{\rm obs}$ 
derived from 1000 realizations of CMB anisotropy sky maps with
$n_{\rm in}=1.25$.
The dotted line is for an initial training
range of $n^{(0)}=0.5$ and $n^{(1)}=1.5$
while the solid line is the distribution for the final
range of $n^{(0)}=0.8$ and $n^{(1)}=1.4$.
The fitted values correctly peak at the input value
(vertical solid line),
despite never having trained on this parameter value.
}
\label{param_hist}
\end{figure}

We generate simulated COBE maps using sky pixels of size 
$2.5^\deg\!\!\times\!2.5\deg$ each. After the galaxy cut, this leaves an input
pattern of $\Ninput=3881$ pixels.  We use $N_{\rm Hidden} = 600$ and $N_{\rm
Train}=12000$  (determined to be the optimal choice) and train 50 networks over
the  parameter range $n^{(0)}=0.0$ and $n^{(1)}=2.0$.  Figure
\ref{param_hist} plots the probability distribution of $n_{\rm obs}$ for a set
of 1000 samples of $n_{\rm in}=1.40$ (dotted line). 
This  distribution matches with separation of the training
sets and we need this only training iteration.
We recover $\bar n_{\rm obs}=1.30$, with 68\% of the samples between
$1.07$ and $1.51$.
In terms of our Bayesian analysis, for $n\sim 1$, we determine
 $\sigma = 0.35$, in agreement with the traditional maximum
likelihood analysis of the COBE-DMR 2 year data \cite{Gorski94b}.

Note that the neural network algorithm recovered the correct spectral index
even though none of the networks used were trained at this value. The
uncertainty, derived from the width of the probability distribution of 
$n_{\rm obs}$, is comparable to the value predicted by 
the maximum likelihood method.
Neural networks can recover cosmological parameters from CMB data sets with
comparable precision as maximum likelihood techniques, but using $N^{1.5}$
calculations instead of $N^3$.


\subsection{Large-Scale Structure}

Large-scale structure provides a third example that combines  parameter
estimation and  phase features. Redshift surveys such as the 2-Degree Field and
the Sloan Digital Sky Survey measure the redshift and position on the sky of a
large number of galaxies ($N \sim 10^6$), sampling the quasi-linear regime
$~\sim 100 h^{-1}$ Mpc where $h$ is the Hubble constant in units 
100 km s${}^{-1}$ Mpc${}^{-1}$.
The observed redshift is the sum of the Hubble flow and the peculiar velocity
induced by gravitational acceleration in the evolving density field. Coherent
flows on large scales  produce artifacts in the redshift distribution compared
to real space. Galaxies on the far side of an overdensity tend to flow toward
the center (hence toward the observer) so that their peculiar velocities
subtract from the Hubble flow, making them appear closer than they really are.
Galaxies on the near side move the opposite direction, so their peculiar
velocities add to the Hubble flow. The net result is an apparent enhancement 
in the galaxy density in redshift space on scales of superclusters, compressing
the region along the line of sight to the observer. The amplitude of this
``bull's-eye'' effect  depends on the matter density $\Omega_m$ on scales
comparable to superclusters of galaxies and can be used to determine $\Omega_m$
in model-independent fashion \cite{Praton97,Melott98}.

Estimating $\Omega_m$ from distortions in redshift space has several problems
in practice. The first is defining a statistic to quantify the bull's-eye
enhancement in concentric rings about the origin. \cite{Melott98} use a large
number of simulations to develop an empirical statistic defined as the ratio of
rms spacing between upcrossings in isodensity contours in the redshift (radial)
direction  to that in the orthogonal (azimuthal) direction. It is thus a local
statistic in that it compares high-density regions only to other nearby
regions, and operates only on a single slice of redshift space after smoothing
and contouring.

\begin{figure}
\plotone{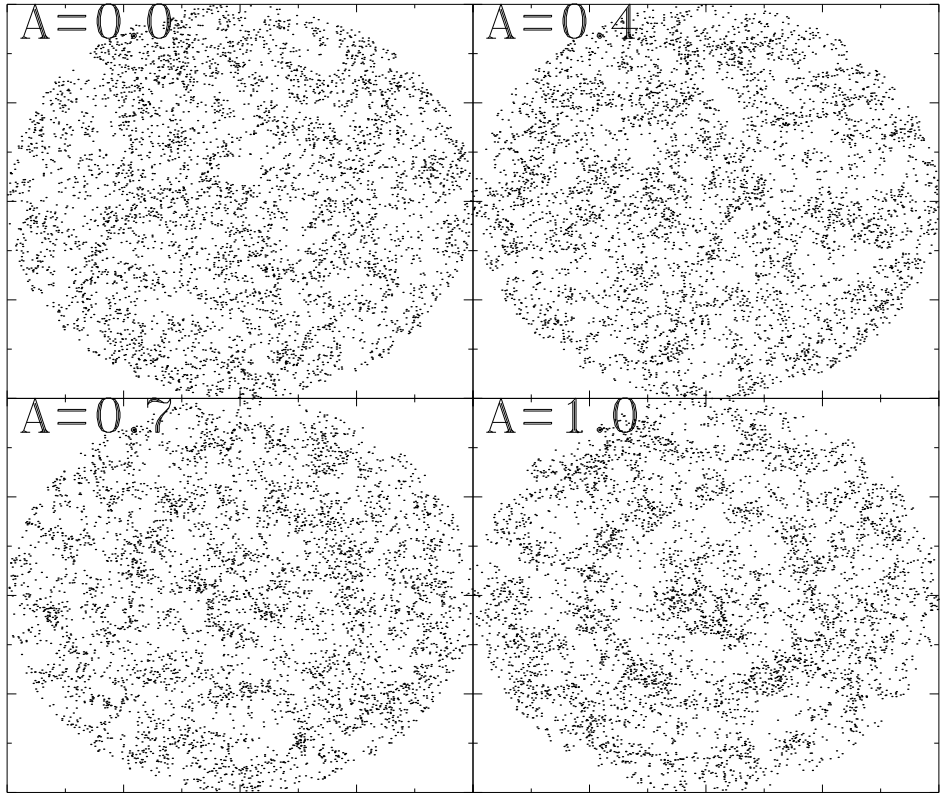}
\caption{
Toy models showing changes in the density enhancement $A$ in redshift space 
due to peculiar velocities (the ``bull's-eye effect''). The same neural nets 
developed for CMB sky maps successfully discriminate at 90\% confidence
among all models shown.
}
\label{bullseye_fig}
\end{figure}

Neural nets, by contrast, offer a {\it global} test by comparing each region of
the density field to all other regions simultaneously, and can easily be
extended across the entire three-dimensional survey. No {\it a priori}
statistic need be identified, nor do neural nets require contouring of the
density field, thus avoiding the need to ``fine-tune' the selection of contour
levels.  Figure \ref{bullseye_fig} shows a toy model of the bullseye effect for
several samples differing in the amplitude of the radial density
enhancements. Our toy model consists of using a Bessel function to modify
an otherwise uniform distribution of galaxies, with the amplitude 
determining the
amount of modulation. The inputs to the neural network are density functions,
one for each realization.
Neural nets, trained to realizations differing only slightly in the amount
of modulation, correctly discriminate between them. 
We do not specify how to tell them apart, but nonetheless,
the correct amplitudes are recovered 90\% of the time. 
Note that we test the bullseye
effect using the {\it same} neural nets developed for 
the noisy sine wave and CMB tests above,
simply trained to a different set of reference models. Neural nets are a
powerful, versatile tool for a variety of problems in astrophysics.


\section{Computational Cost Scaling}

Neural nets have many desirable characteristics for parameter estimation with
mega-pixel CMB maps. They operate globally on the data and return unbiased
estimates of the underlying parameter values. They automatically account for
data gaps, instrument noise, and other features peculiar to a particular data
set. Most importantly, the computational costs are low enough to allow
extension to the mega-pixel data sets expected in the near future.

The dominant contribution to the CPU cost is the array multiplication 
associated with the weights.  This multiplication is performed twice per
training pass: once for evaluating the pattern, then again for back propagating
the error.  Each operation scales as the number of weights.  Our total
computational cost for training a network is 
\beq
N_{\rm CPU} = 2\; N_{\rm Train} N_{\rm Hidden}\left(\Ninput+1\right)
\eeq
How this cost scales with $\Ninput$ depends on the problem being considered. We
find that $N_{\rm CPU} \sim N^\alpha$ with $1< \alpha < 2.5.$ For the
particular application of mega-pixel CMB maps, $N_{\rm CPU} \sim N^{1.5}_{\rm
Input}$.

The computational cost depends on  how the information content of the input
pattern changes as more data points are added. We consider two limiting cases.
The first case has fixed signal to noise per data point, so that the signal to
noise ratio of the entire data set improves as more points are added. This is
analogous to measuring a signal with a noisy detector, and simply adding more
observations. The second case keeps the signal to noise ratio of the entire
data set fixed, so that the signal to noise ratio per pixel gets worse  as more
pixels are added. This is analogous to over-sampling a signal within a limited
observing time: as the observations are broken into finer and finer resolution,
the observing time per point decreases and the noise per observation gets
worse.

We treat these limiting cases using our noisy sine model.  For the first case
(fixed noise per observation), we simply simulate more data over additional
periods of the sine wave, $t_f$ grows with $\Ninput$ (maintaining both the
pattern of regular cutouts  and the percentage of random cutouts). The noise
per pixel is fixed at  $\sigma_i^{\rm noise} = 1$, independent of $\Ninput$.
For the second case (fixed noise averaged over the entire data set), we keep
the number of periods per data set fixed, $t_f = 1$, but allow the noise
amplitude to vary as  $\sigma_i^{\rm noise} = \sigma_0/\sqrt{N^{\rm obs}_i}$,
and fix $\sigma_0$  so $\sigma_i^{\rm noise} = 1$ when $\Ninput = 256$.

\begin{figure}
\plotone{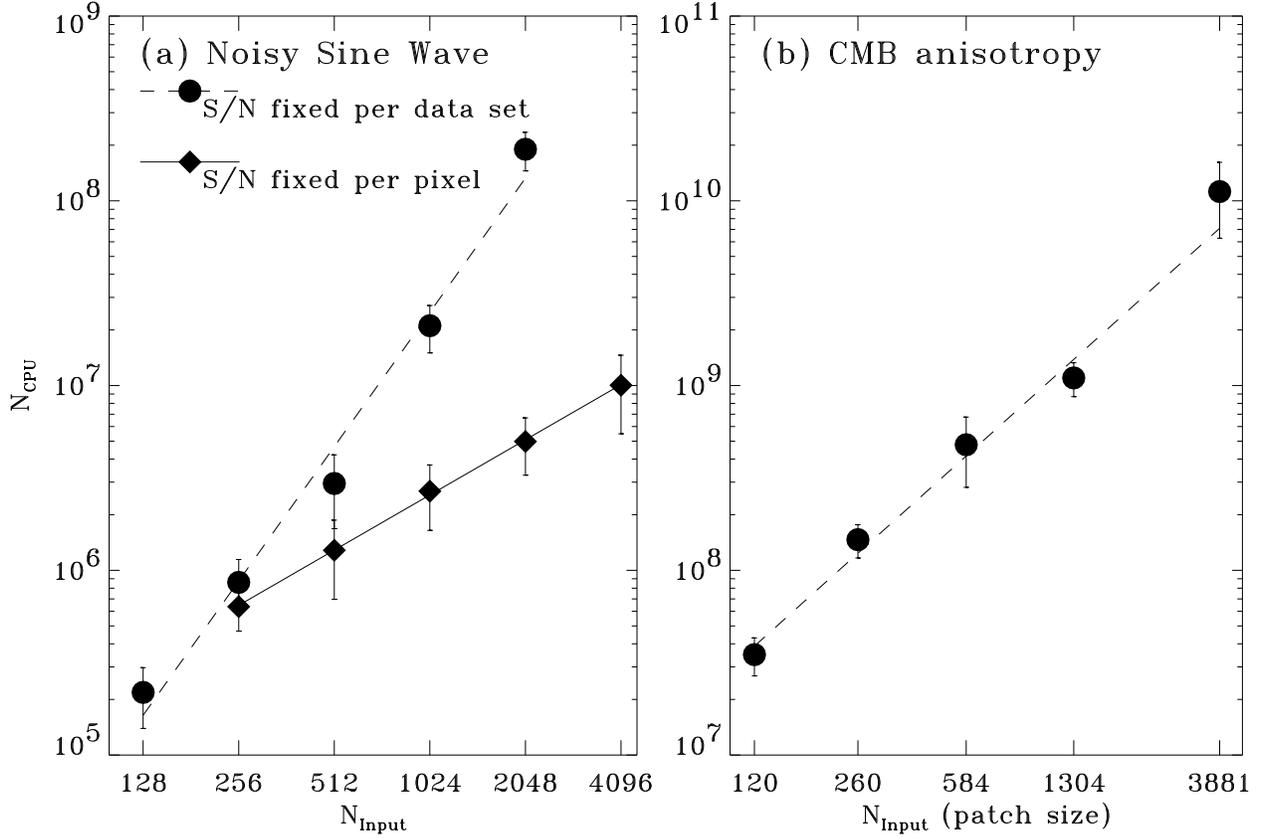}
\caption{
CPU cost scaling
(a) Scaling of computational costs for the limiting cases
of irregularly sampled noisy sine wave. 
We consider two cases: 
i) the S/N per pixel is fixed (diamonds/solid line); 
and ii)
the S/N for the entire pattern is fixed and
the noise per pixel grows with $\Ninput$ (circles/dashed line).
These limiting cases are well described by power-law fits
$N_{\rm CPU}\sim N^{\alpha}_{\rm Input}$
with $\alpha$ = 1.0 and 2.5, respectively.
(b) Scaling of computational costs for CMB anisotropy. Working at a fixed
sky map resolution, we vary the patch size that is examined. This holds
the S/N per pixel fixed, but new information is introduced as the patch
size increases. The solid line represents a power law fit of
$N_{\rm CPU} \sim \Ninput^{1.5}$.
}
\label{ncpu}
\end{figure}

The number of training passes and the number of hidden units depend on problem
considered and the number of input pixels. For each case above, we determine
which values give the best $N_{\rm CPU}$ for a fixed training accuracy. 
For a range of $N_{\rm Hidden}$, we
train multiple networks  on a large set of patterns.  As they are being
trained, we monitor their ability  to correctly classify independent samples. 
Once this ability passes a pre-set threshold,  we know $N_{\rm Train}$ for each
network, and hence $N_{\rm CPU}$.  We repeat for multiple network to estimate
the uncertainty in the CPU cost. (We may also vary other training parameters to
assure we have the optimal results.)  The results are not strongly  dependent
on the precise value of $N_{\rm Hidden}$.  Above a minimum value, as $N_{\rm
Hidden}$ increases,  the number of training passes needed decreases in such a
way that  $N_{\rm CPU}$ remains constant over a wide range of $N_{\rm
Hidden}$.  For these limiting cases, we find $N_{\rm CPU} \sim \Ninput$
for fixed noise per pixel, and $N_{\rm CPU}\sim \Ninput^{2.5}$
for fixed noise per data set (Figure \ref{ncpu}(a)).

The computational costs for CMB analyses lie between these extremes. We derive
the scaling for CMB maps by analyzing the CPU costs as progressively larger and
larger areas of the sky are covered. In this way, new information is introduced
into the data sets as the patch size increases. The S/N ratio per pixel is
fixed in this scheme, reflecting the trend in current experiments of scanning
ever larger portions of the sky at (roughly) constant S/N ratio per pixel.

We select circular patches of sky centered at the north zenith.  The range of
patch sizes are chosen to cover 1.5 orders of magnitude in $\Ninput$. For each
patch size, we proceed as above to determine the optimal CPU cost. The results
are shown in Figure \ref{ncpu}(b). The computational cost for training a CMB
network scales according to $N_{\rm CPU}\sim \Ninput^{1.5}$, a
considerable improvement over the $\Ninput^3$ scaling behavior for a
maximum likelihood analysis.


\section{Discussion}

We have shown neural networks can be used as a tool for astrophysical
parameter estimation. For specificity, we have worked in a cosmological context
(CMB maps and redshift surveys) where the stochastic nature of the problem
is fundamental. The
results are insensitive  to noise levels and sampling schemes typical of large
astrophysical data sets and provide parameter estimation
comparable to maximum likelihood techniques. The computational cost is never
greater than standard maximum likelihood techniques and in the context
for CMB anisotropy maps, we find $N_{\rm CPU} = O(\Ninput^{1.5})$.


If we classify parameter estimation techniques as to whether they
are forward or reverse algorithms, we see the real strength of neural
networks.
Maximum-likelihood methods are an
example of reverse algorithms.
They start with the statistic under consideration
and work backwards, inverting a covariance matrix, 
to the likelihood function
used to compare different parameter choices.
Forward algorithms provide a way to avoid the high computational costs of
inverse methods. Typically, it is much simpler to generate model predictions 
at each sampled point in parameter space  than to compute the matrix inverse
and determinant
required for  maximum likelihood techniques. Forward algorithms trade many
realizations of synthetic data sets computed at specific parameter values for
the computationally infeasible matrix inversion. Neural networks are such an
algorithm; synthetic data sets are used to both train and sample the
networks. This gives us our speed improvement.

\begin{figure}
\plotone{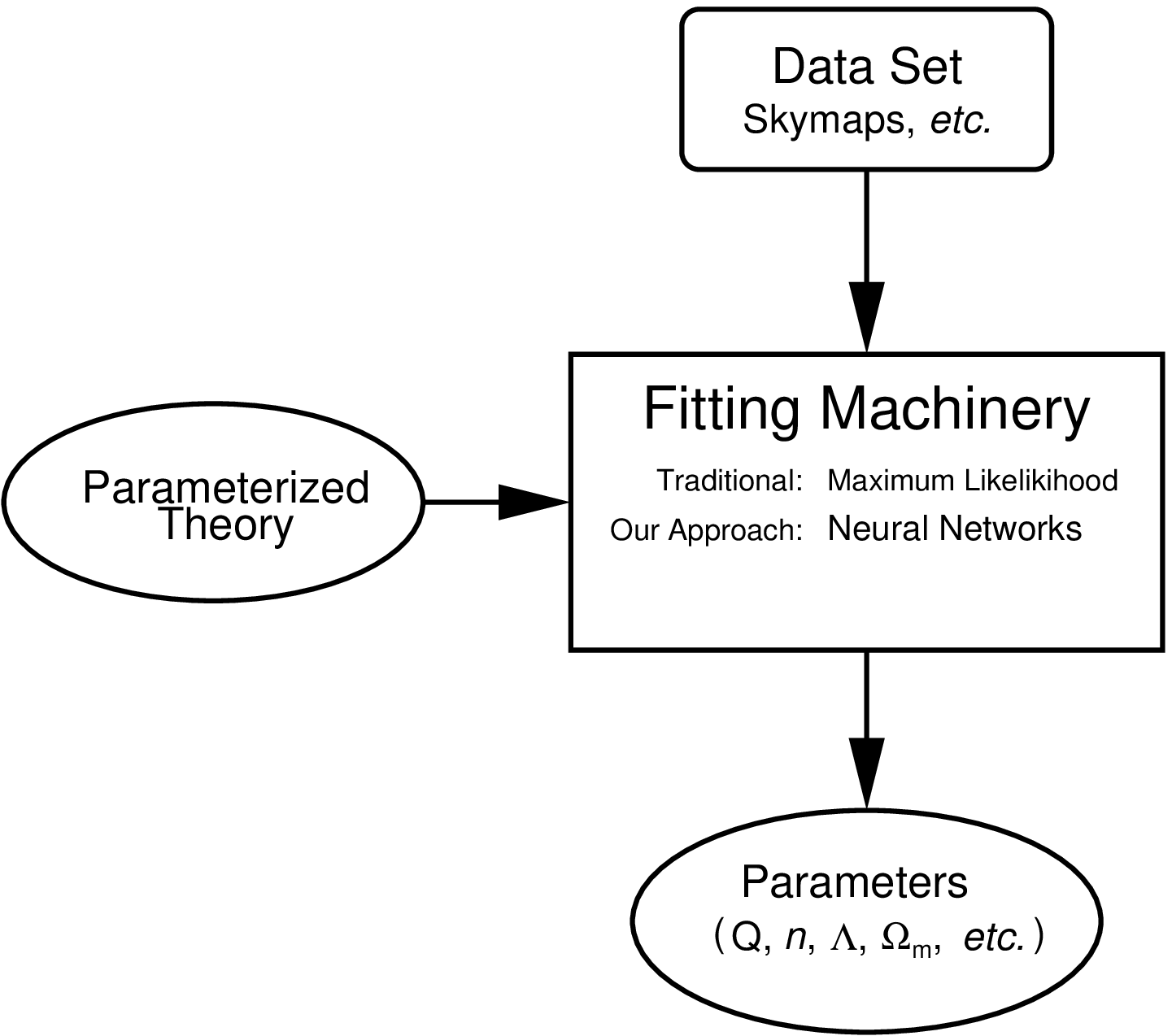}
\caption{
Our basic model for parameter estimation from a data set. We have to {\it a
priori} assume a model to compare the observed data set against; what we pay
attention to is the machinery for performing this comparison. Maximum
likelihood methods, the {\it de facto} standard in the CMB community must
assume a model, just as must be done with currently proposed neural network
method.
}
\label{fitting-model}
\end{figure}

Since either maximum likelihood or neural networks can be viewed as the
``machinery'' for parameter estimation, the fundamental information
flow stays the same (see figure \ref{fitting-model}). The statistical
confidence levels for the fitted parameters are always accessible. When
the ``machinery'' is sampled with independent  synthetic data, we can
determine the probabilities for making correct or incorrect parameter
identifications. Such sampling also gives us direct access to the
statistical power \cite{PhillipsKogut01}. While training, the information
distinguishing the different parameters is encoded in the weights.
Interperting the resulting weight matrices is not usually possible
(as compared to the Fisher matrix, Eqn \ref{fisher_def}). Using 
independent sampling
of the network to derive the probability distributions needed for,
{\it e.g.} Bayesian analysis, means we do not need direct access to the
information in the weight matrices.

A limitation of
maximum likelihood methods is their
requirement of an {\it a priori} definition of  a goodness-of-fit
function.
The choice of a goodness-of-fit function is not always  obvious and
is particularly acute for 2D and 3D surveys.
Much of the information lies in the {\it phase} features of these surveys.
Statistical tests can fail badly in detecting phase features, as witness the
large literature devoted to the relatively simple problem of edge detection in
2D data sets (see, e.g., \cite{Hough62,Davis75,Canny86}). Topological tests
such as the genus or other Minkowski functionals have been applied to 2- and
3-D maps, \cite{Gott90,Kogut96}
but the relative power of these statistics is poor
\cite{PhillipsKogut01}. 
Neural networks, in contrast, do not require specification of
a single statistic of {\it a priori} interest.
As the network is trained, it determines how it will discriminate
between competing models.
The information required to separate different parameter values
comes from the training set simulations.
Neural networks thus offer a promising approach to
cosmological parameter estimation, 
where the statistical properties of the primordial matter and energy
distribution provide one of the few falsifiable tests  of the standard
inflationary paradigm.


\newpage
%

%

\begin{thebibliography}{}

\bibitem[Andreon \etal\ 2000]{Andreon00}
Andreon, S., \etal\ 2000, \mnras, in press

\bibitem[Bailer-Jones \etal\ 1997]{BailerJones97}
Bailer-Jones, C. A. L., Irwin, M., Gilmore, G. and von Tippel, T. 1997, 
\mnras, 292, 157

\bibitem[Bailer-Jones \etal\ 1998]{BailerJones98}
Bailer-Jones, C. A. L., Irwin, M. and von Tippel, T. 1998, 
\mnras, 298, 361

\bibitem[Bailer-Jones \ 2000]{BailerJones00}
Bailer-Jones, C. A. L. 2000, Astron. Astrophys., 357, 197


\bibitem[Banday \etal\ 1997]{Banday97} 
Banday, A.J.,
G\'orski, K.M., Bennett, C.L., Hinshaw, G., Kogut, A.,
Lineweaver, C.,  Smoot, G.F., and Tenorio,  L. (1997) ApJ, 475, 393

\bibitem[Bennett \etal\ 1995]{Bennett95}
Bennett, C.L., et al.\ 1995, BAAS, 187, 7109;
http://map.gsfc.nasa.gov/

\bibitem[Bennett \etal\ 1996]{Bennett96}
Bennett, C.L., et al.\ 1996, ApJ, 464, L1

\bibitem[Bersanelli \etal\ 1996]{Bersanelli96}
Bersanelli, M., et al.\ 1996, {\it COBRAS/SAMBA},
ESA Report D/SCI(96)3;
http://astro.estec.esa.nl/Planck/

\bibitem[Bishop\ 1995]{Bishop95}
Bishop, C.\ 1995, {\it Neural Network for Pattern Recognition} 
(Oxford University Press)

\bibitem[Bond and Efstathiou\ 1987]{Bond87}
Bond, J.\ R., and Efstathiou, G.\ 1987, MNRAS, 226, 655

\bibitem[Bond \etal\ 1997]{Bond97}
Bond, J.\ R., Efstathiou, G., and Tegmark, M.\ 1997, MNRAS, 291, L33

\bibitem[Bond \etal\ 2000]{Bond00}
Bond, J.\ R., et al.\ 2000, in
Proc. IAU Symp. 201, preprint astro-ph/0011378

\bibitem[Borrill\ 1999]{Borrill99}
Borrill, J.\ 1999, PRD, 59, 027302


\bibitem[Canny\ 1986]{Canny86}
Canny, J.\ 1986, IEEE Trans. Pattern Analysis and Machine Intelligence,
8, 679

\bibitem[Charbonneau\ 1995]{Charbonneau95}
Charbonneau, P.\ 1995, ApJS, 101, 309

\bibitem[Christensen and Meyer\ 2000]{Christensen00}
Christensen, N. and Meyer, R. \ 2000, {\tt astro-ph/0006401}

\bibitem[Copi \etal\ 1995]{Copi95}
Copi, C.J., Schramm, D. N., \& Turner, M. S. 1995, Science, 267, 192


\bibitem[Davis\ 1975]{Davis75}
Davis, L.\ S.\ 1975, 
Computer Graphics and Image Processing, 4, 248

\bibitem[deBernadis \etal\ 2000]{deBernadis00}
deBernadis, P., \etal\ 2000, Nature, 404, 955

\bibitem[Dor\'e \etal\ 2001]{Dore01}
Dor\'e, O., Knox, L. and Peel, A., preprint astro-ph/0104443

\bibitem[Douspis \etal\ 2001]{Douspis01}
Douspis, M., Bartlett, J. G., Blanchard, A. and Le Dour, M.,
preprint astro-ph/0104366

\bibitem[Ferreira \etal\ 1998]{Ferreira98}
Ferreira, P.G., Magueijo, J., \& G\'{o}rski, K.M.\ 1998, ApJL, 503, L1

\bibitem[Goldberg\ 1989]{Goldberg89}
Goldberg, D.\ E.\ 1989,
{\it Genetic Algorithms in Search, Optimization, and Machine Learning}
(Reading, MA: Addison-Wesley)


\bibitem[G\'{o}rski \etal\ 1994]{Gorski94b}
G\'{o}rski, K.\ M., 1994, ApJL, 430, L89

\bibitem[G\'{o}rski \etal\ 1996]{Gorski96}
G\'{o}rski, K.\ M., et al.\ 1996, ApJL, 464, L11


\bibitem[G\'{o}rski \etal\ 1999]{Gorski99}
G\'{o}rski, K.\ M.,  Wandelt, B.\ D., Hansen, F.\ K., Hivon, E., and
Banday, A.\ J.\ 1999, preprint astro-ph/9905275 {\tt published?}

\bibitem[Gott \etal\ 1990]{Gott90}
Gott, J.R., et al.\ 1990, ApJ, 352

\bibitem[Grossberg\ 1988]{Grossberg88}
Grossberg, S.\ 1988, Neural Networks, 1, 17


\bibitem[Hassibi and Stork\ 1993]{Hassibi93}
Hassibi, B. and Stork, D. G. 1993, in {\it Advances in Neural Informatiom Processing
Ssytems}, Volume 5, 164 (Morgan Kaufman:San Mateo, CA)

\bibitem[Hinshaw \etal\ 1994]{Hinshaw94}
Hinshaw, G., \etal\ 1994, ApJ, 431, 1

\bibitem[Hinshaw \etal\ 1995]{Hinshaw95}
Hinshaw, G., \etal\ 1995, ApJL, 446, L67

\bibitem[Hinshaw \etal\ 1996a]{Hinshaw96a}
Hinshaw, G., \etal\ 1996, ApJL, 464, L17

\bibitem[Hinshaw \etal\ 1996b]{Hinshaw96b}
Hinshaw, G., \etal\ 1996, ApJL, 464, L25

\bibitem[Hivon \etal\ 2001]{Hivon01}
Hivon, E., G\'orski, K. M., Netterfield, C. B., Crill, B. P.,
Prunet, S. and Hansen, F., preprint astro-ph/0105302

\bibitem[Holland\ 1975]{Holland75}
Holland, J.\ H.\ 1975,
{\it Adaptation in Natural and Artificial Systems}
(1st ed.; Ann Arbor: Univ. of Michigan Press;
2nd ed.; 1992, Cambridge: MIT Press)

\bibitem[Hopfield\ 1982]{Hopfield82}
Hopfield, J.J.\ 1982,
Proceedings of the National Academy of Sciences of the USA, 79, 2554

\bibitem[Hough\ 1962]{Hough62}
Hough, P.V.C., 1962, 
{\it Method and Means for Recognizing Complex Patterns}, 
US Patent 3,069,654

\bibitem[Hu and Sugiyama\ 1995]{HuSugiyama95}
Hu, W. and Sugiyama, N.\ 1995, PRD, 51, 2599

\bibitem[Hu 1999]{Hu99}
Hu, W. in {\it Birth and Evolution of the Universe} RESCEU {\it 1999};
preprint astro-ph/0002520


\bibitem[Kogut \etal\ 1995]{Kogut95}
Kogut, A., \etal\ 1995, ApJ, 439, L29

\bibitem[Kogut \etal\ 1996]{Kogut96}
Kogut, A., \etal\ 1996, ApJL, 464, L29


\bibitem[Lahav \etal\ 1996]{Lahav96}
Lahav, O., \etal\ 1996, \mnras, 283, 207


\bibitem[Le Cun \etal\ 1990]{LeCun90}
Le Cun, Y., Denker, J. S. and Solla, S. A.\ 1990, in {\it Advances in Neural 
Informatiom Processing Ssytems}, Volume 2, 598 (Morgan Kaufman:San Mateo, CA)

\bibitem[MacKay 1995]{MacKay95}
MacKay, D. J. C.\ 1995,
Network: Computation in Neural Systems, 6 469

\bibitem[McCulloch  and Pitts\ 1943]{McCulloch43}
McCulloch, W., and Pitts, W. 1943,
Bulletin of Mathematical Biophysics, 7, 115

\bibitem[Melchiorri \ 2000]{Melchiorri00}
Melchiorri, A., et al.\ 2000, ApJ, 536, L63

\bibitem[Melott \etal\ 1998]{Melott98}
Melott, A.\ L., Coles, P., Feldman, H., and Wilhite, B,\ 1998, 
ApJL, 496, L85

\bibitem[Netterfield \etal\ 2001]{Netterfield01}
Netterfield, C. B., \etal, submitted to ApJ, preprint astro-ph/0104460

\bibitem[Oh \etal\ 1999]{Oh99}
Oh, S.P., Spergel, D.N., and Hinshaw, G.\ 1999, ApJ, 510, 551

\bibitem[Olive \etal\ 2000]{Olive00}
Olive, K. A., Steigman, G. \& Walker, T. P. 2000, Phys. Rep., 333, 389

\bibitem[Pando \etal\ 1998]{Pando98}
Pando, J., Valls-Gabaud, D, and Fang, L.-Z.\ 1998, PRL, 81, 4568

\bibitem[Peebles and Yu \ 1970]{PeeblesYu70}
Peebles, P. J. E. and Yu, J. T.\ 1970, ApJ, 162, 815

\bibitem[Phillips and Kogut\ 2001]{PhillipsKogut01}
Phillips, N.G., and Kogut, A.\ 2001, ApJ, 548, 540

\bibitem[Praton \etal\ 1997]{Praton97}
Praton, E.\ A., Melott, A.\ L., and McKee, M.\ Q.\ 1997, ApJL, 479, L15


\bibitem[Rocha\ 2000]{Rocha00}
Rocha, G., \etal\ 2000, \mnras, in press

\bibitem[Rosenblatt\ 1962]{Rosenblatt62}
Rosenblatt, F.\ 1962,
{\it Principles of Neurodynamics} (Spartan Books)

\bibitem[Rumelhart \etal\ 1986a]{Rumelhart86a}
Rumelhart, D.E., Hinton,  G.E., and McClellend, J.L.\ 1986,
in {\it Parallel Distributed Processing}, 
Eds. D.E.\ Rumelhart, J.L.\ McClellend and the PDP Research Group 
(MIT Press: Cambridge)

\bibitem[Rumelhart \etal\ 1986b]{Rumelhart86b}
Rumelhart, D.E., Hinton, G.E.,  and Williams, R.J.\ 1986,
in {\it Parallel Distributed Processing}, 
Eds. D.E.\ Rumelhart, J.L.\ McClellend and the PDP Research Group 
(MIT Press: Cambridge)

\bibitem[Seljak and Zaldarriaga\ 1996]{Seljak96}
Seljak, U., and Zaldarriaga, M. 1996, ApJ, 469, 437

\bibitem[Szapudi \etal\ 2000]{Szapudi00}
Szapidu, I., Prunet, O., Pogosyan D., Szalay, A. S. and Bond, J. R., 
preprint astro-ph/0010256

\bibitem[Tagliaferri \etal\ 2000]{Tagliaferri00}
Tagliaferri, R., \etal\ 2000 \aap, in press




\bibitem[Watanabe\ 1969]{Watanabe69}
Watanabe, S.\ 1969,
{\it Methodologies of Pattern Recognition}
(Academic Press: New York)

\end{thebibliography}
\end{document}